# *Rational Doping Strategy of Porous Materials for Hydrogen Storage: CNTs study case*

Shima Rezaie [a], Azahara Luna-Triguero,*[a,b]

[a] *Energy Technology, Department of Mechanical Engineering, Eindhoven University of Technology, P.O. Box 513, 5600 MB Eindhoven, The Netherlands.*

[b] *Eindhoven Institute for Renewable Energy Systems (EIRES), Eindhoven University of Technology, PO Box 513, Eindhoven 5600 MB, The Netherlands*

**Abstract**

Identifying a nanostructure suitable for hydrogen storage presents a promising avenue for the secure and cost-effective utilization of hydrogen as a green energy source. This study introduces a systematic approach for selecting optimal doping on porous materials, emphasizing the intricate interplay between doping with the material's structure and the interaction between doping and hydrogen. Our proposed approach serves as a framework for evaluating and predicting the performance of doped materials. To validate the efficacy of our strategy, we conduct a comprehensive investigation in carbon nanotubes (CNTs). Applying our criteria, we systematically screen several dopants in CNTs. The results highlight Cu-doped CNTs as promising candidates for hydrogen storage applications. Focusing on Cu-doped CNTs, we analyze binding energy, charge transfer, partial density of states (PDOS), and desorption temperature to assess the performance of modified CNTs. Additionally, we explore the feasibility of doped CNTs featuring various sizes of copper clusters and the effect on the release temperature, i.e., complete regeneration. The findings indicate that incorporating 5 to 6% copper impurity onto CNT surfaces renders these nanostructures highly applicable for reversible hydrogen storage near ambient conditions.

## Introduction

The importance of energy in modern life is undeniable, as it underprints all kinds of activities, including transportation, manufacturing, space heating or cooling, cooking, lighting, among others [1], [2]. The primary source of this energy has historically been fossil fuels, including oil, petroleum, gas, and coal [2]. However, the excessive reliance on these finite resources has resulted in rapid depletion and detrimental environmental and health consequences due to carbon dioxide emissions [3]. Therefore, there is an urgent need to identify a clean and sustainable energy source to meet the growing global energy demand [4].

Energy production based on burning hydrogen gas is the most promising solution to solve the problem of energy demand. As the most abundant element on earth, hydrogen has the highest energy density per mass compared to all other energy sources (energy density about 120 MJ/kg) [5]. Hydrogen fuel cells have achieved an impressive maximum efficiency of around 65%, significantly surpassing traditional fuels like diesel (45%) and gasoline (22%) [6], [7]. The density of hydrogen is 3.2 times lower than natural gas and 2700 times lower than gasoline [6]. Moreover, hydrogen combustion produces as by-products water vapor and heat, making it an environmentally friendly option with no greenhouse gas emissions [8]. These significant features position hydrogen gas as a frontrunner for a sustainable energy future [8], [9].

Despite its potential, the practical utilization of hydrogen as a fuel source hinges on effective hydrogen storage. Storage enables energy to be available when needed, ensuring a consistent supply that complements other renewable sources to mitigate fluctuations in energy production due to varying weather conditions and seasons [10], [11]. As an example, excess energy generated during summer days can be used to produce and store hydrogen, which can be tapped during winter when solar energy production is limited [11]. In this scenario, hydrogen storage is paramount for using hydrogen as fuel.

Hydrogen is a light and flammable gas with low density in ambient conditions [12], [13]. Therefore, the storage process should be carried out at high pressures and low temperatures to ensure a practical energy supply. Common hydrogen storage methods: *i*) compressed gas; which occurs at high pressures (minimum pressure started at 700 atm; 10000 psi), *ii*) cryogenic liquid (20K; -253 ºC); which occurs at very low temperatures, and *iii*) solid-state material; which can happen at different pressures and temperatures depending on the material [14]. Providing proper storage conditions for the first and second methods is not cheap, especially when it comes to safety issues for portable applications [15]–[17]. In this regard, storing hydrogen in solid-state materials can be the most efficient way in terms of safety and cost-effective solution to avoid using hydrogen at extremely low temperatures and high pressures [15]–[17].

In general, hydrogen storage in nanomaterials can be driven by physisorption, chemisorption, and capillary condensation mechanisms [9]. In chemisorption, hydrogen molecules are typically dissociated and integrated into the material [9]. Although this method can lead to high hydrogen capacity, irreversibility remains the main drawback. Capillary condensation is a phenomenon that causes gases in porous materials to liquefy at low pressures. This mechanism can lead to high volumetric capacity in storing molecular hydrogen, but the network of the nanomaterial must be carefully controlled. Physisorption, based on weak van der Waals interactions between hydrogen and the material surface, can provide a reversible hydrogen storage with safety and low cost [18], [19].

The critical challenge in using solid hydrogen storage lies in identifying a suitable material capable of reversibly storing hydrogen. According to the Department of Energy of the US (DOE) standards, efficient materials for hydrogen storage applications should exhibit gravimetric capacity with a lower limit of about 5.5 wt% by 2025 and an ultimate goal 6.5 wt% [20], [21]. In addition, the binding energy should be in the range of 0.15 to 0.6 eV for reversible hydrogen storage [14].

Among all the discovered nanomaterials, carbon nanotubes (CNTs) have been widely used for different applications ranging from energy storage, sensors, gas storage, biomedical, and optoelectronics [18]–[24]. The most important features of CNTs are large surface area, high stability, wide diversities of pore structure, ability to absorb gases at ambient temperature, tunable properties, and lightweight nature. These properties make them suitable materials for hydrogen storage purposes [25], [26]. Despite these advantages, the average binding energy between hydrogen and the surface of CNT is typically low, resulting in limited storage capacity at room temperature and pressure [27], [28]. Experimental and theoretical studies have consistently reported that introducing dopants to the surface of CNTs significantly enhances the binding energy between hydrogen and the CNT surface [28]–[40]. However, the pivotal question remains: among the numerous potential dopants, which can provide reversible hydrogen adsorption with high gravimetric capacity in CNTs?

To address this question, simulations offer a powerful and cost-effective approach to test the performance of a wide range of materials and providing informed-decision guidelines for experiments. In this context, density functional theory (DFT) has been used as a valuable method to extract crucial insights.

In the present study, we propose a transferable approach to select the dopant of a material for hydrogen storage by using binding energies as critical criteria. To test the strategy, 29 impurities, spanning all groups of the periodic table, have been tested for potential surface addition on the external surface of CNTs. Among all these impurities, intriguing results have been obtained by adding copper to the surface of CNTs. Simulations have been performed on three different CNTs to provide comprehensive insights. We explored the effect of CNT diameter and the number of layers. Binding energy, charge transfer, PDOS, DOS, and desorption temperature profiles have been analyzed to assess the suitability of Cu-doped CNTs for hydrogen storage applications. Furthermore, the effect of copper clustering on the adsorption capability and the desorption temperature of these structures has been evaluated.

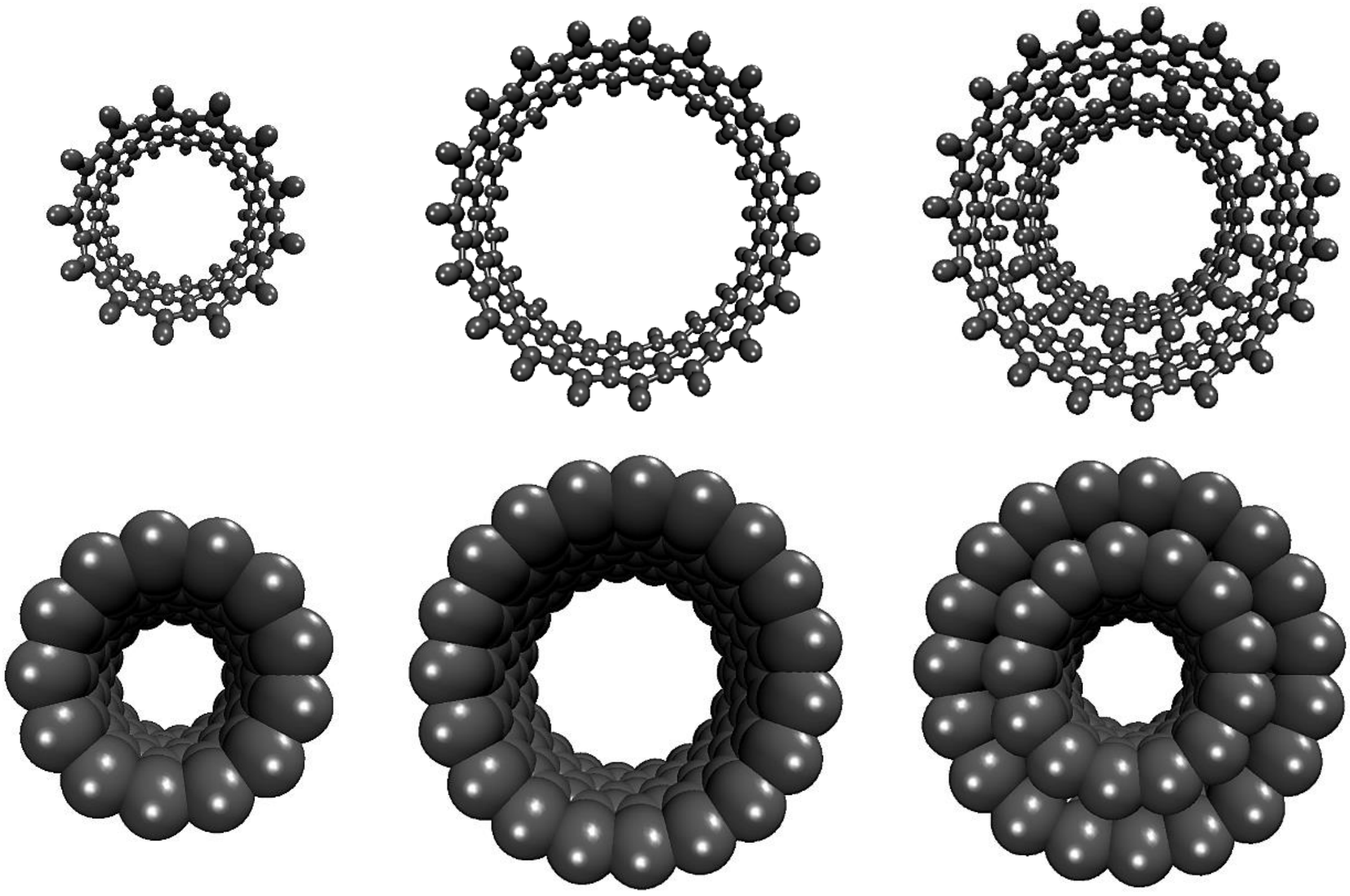

*Figure 1. Schematic representation of three zigzag CNT nanostructures.*

## Calculation methodology

The hydrogen storage performance of Cu-doped CNTs has been performed using Density Functional Theory (DFT) implemented in the SIESTA package [41]. The exchange-correlation energy has been described using the generalized gradient approximation (GGA) developed by Perdew-Burke-Ernzerhof (PBE) [42]. To describe the potential of atomic cores and related core electrons with valence electrons the Troullier-Martins pseudopotentials have been used. The expansion of the wave functions at the double-zeta plus polarization (DZP) level has occurred on numerical atomic orbitals (NAOs). A force tolerance of 0.04 eV/Å has been considered for all relaxation coordinates. The Brillouin zone has been sampled using a 2×2×4 Monkhorst-Pack k-point. In addition, a mesh cut-off energy of 150 Ry has been considered for the plane-wave basis set.

Three different zigzag CNTs have been considered to obtain the required information. The smallest single-walled CNT ($CNT_{(13,0)}$) structure contains 104 carbon atoms per unit cell with a diameter and length of about 10.28 Å and 7.82 Å, respectively. The second single-walled ($CNT_{(19,0)}$) contains 152 carbon atoms per unit cell with a diameter and length of about 15.05 Å and 7.82 Å, respectively. The C-C bond between 1.42 Å to 1.43 Å has been obtained for both $CNT_{(13,0)}$ and $CNT_{(19,0)}$, which is in good agreement with previous studies [28], [43], [44]. The third CNT is a double-walled CNT (DWCNT) made from a combination of the first and second single-walled CNTs (SWCNTs). The DWCNT contains 256 carbon atoms with internal and external diameters of about 9.95 Å and 15.69 Å, respectively. The C-C bond is calculated to be 1.40 Å to 1.41 Å for the internal CNT layer and 1.43 Å to 1.47 Å for the external CNT layer, which is in good agreement with previous studies [45]. The length of the DWCNT is similar to the first and second SWCNTs. Since all structures have the same length, the same lattice parameters of a= 25 Å, b= 25 Å, and c= 8.526 Å have been defined to make infinitely long CNTs. All structures are shown in Figure 1. Moreover, Crystallographic information for all structures can be found in the supplementary material.

The binding energy between copper and the surface of CNTs has been calculated using Eq. (1).

$$E_b(Cu) = E_{tot(CNT)} + E_{tot(Cu)} - E_{tot(Cu-doped\ CNT)} \quad , Eq.\ (1)$$

where $E_{tot(CNT)}$ represents the total energy of the CNT per unit cell, $E_{tot(Cu)}$ indicates the total energy of isolated copper atom, and $E_{tot(Cu-doped\ CNT)}$ shows the total energy of Cu-doped CNT per unit cell.

The average binding energy between hydrogen molecules and the surface of Cu-doped CNT has been computed using Eq. (2).

$$E_b(H_2) = \frac{E_{tot(Cu-doped\ CNT)} + E_{tot(iH_2)} - E_{tot((Cu-doped\ CNT)+iH_2)}}{i} \quad , Eq.\ (2)$$

Where $E_{tot(Cu-doped\ CNT)}$ indicates the total energy of Cu-doped CNT per unit cell, $E_{tot(iH_2)}$ denotes the total energy of *i* isolated hydrogen molecules, $E_{tot((Cu-doped\ CNT)+iH_2)}$ represents the total energy of Cu-doped CNT with *i* adsorbed hydrogen molecules, and i shows the number of adsorbed hydrogen molecules.

For further explanation of the adsorption mechanism of hydrogen molecules on the surface of CNTs, the charge transfer has been calculated by Eq. (3) through the Mulliken charge transfer analysis.

$$Q_t = Q_{(ads-H_2)} - Q_{(iso-H_2)} \quad , Eq.\ (3)$$

$Q_{(ads-H_2)}$ and $Q_{(iso-H_2)}$ represent the total charge of adsorbed and isolated hydrogen molecules, respectively.

The desorption temperature of hydrogen molecules from the surface of Cu-doped CNT has been computed using The Van't Hoff 's equation (Eq. (4) [46]).

$$T_D = \frac{E_{ads}}{k_B}\left(\frac{\Delta S}{R} - \ln p\right)^{-1} \quad , Eq.\ (4)$$

The average adsorption energy, Boltzmann constant, change in hydrogen entropy from gas to liquid phase, gas constant, and atmospheric pressure are represented by $E_{ads}$, $k_B$, $\Delta S$, $R$, and $p$ in Eq. (4).

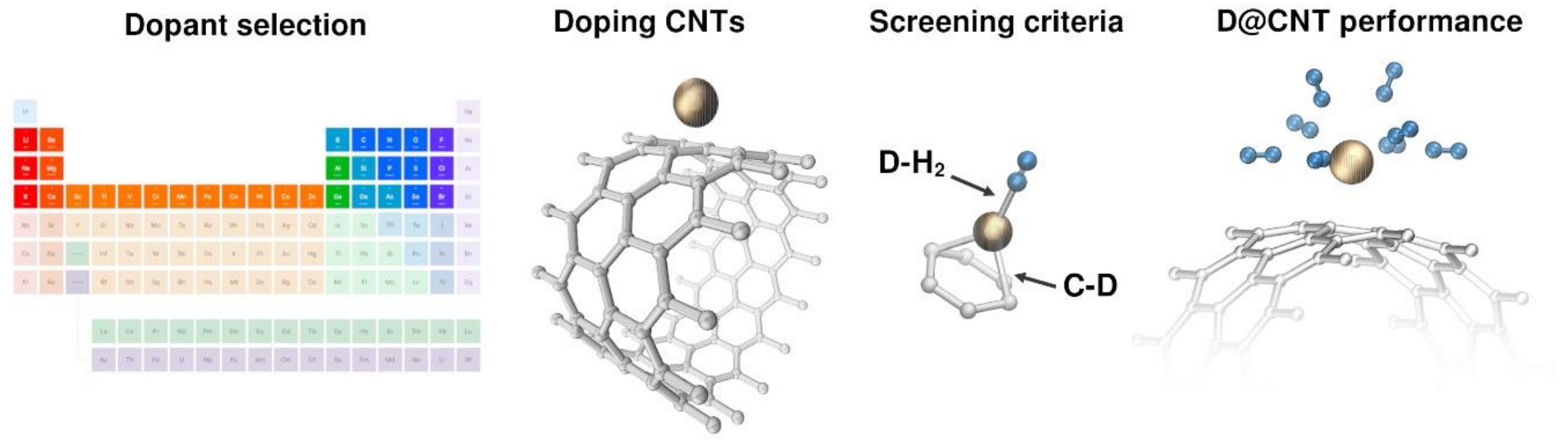

*Figure 2. Schematic representation of the four-step strategy for the doping selection..*

Moreover, the gravimetric capacity is calculated using Eq. (5):

$$\text{wt }\% = \frac{i\,W_{H_2}}{i\,W_{H_2} + W_{(Cu-doped\ CNT)}} \times 100 \quad , Eq. (5)$$

where i represents the number of adsorbed hydrogen molecules, $W_{H_2}$ and $W_{(Cu-doped\ CNT)}$ are the molecular weight of the hydrogen molecule and Cu-doped CNT structure, respectivly. More information on nomenclature and parameters can be found in the Glossary, Table S1 in the Supporting Information (SI).

## Results and discussions

As mentioned earlier, pure CNT do not meet the hydrogen storage expectations set by the DOE. In response, the modification of these structures has been explored as an effective strategy to improve their performance for hydrogen storage applications. In accordance with the delineation in Figure 2, a four-step approach has been systematically employed to identify a suitable impurity for enhancing the hydrogen storage of CNTs. In the first step, prelaminar *dopant selection*, atoms characterized by lower atomic weights are selected to maximize the probability of achieving high gravimetric capacities, almost all periodic table groups (excluding the noble gases) are covered. After the first selection, 29 different impurities are used for *doping CNTs*, adding them to the surface of single and double-walled CNTs and allowing relaxation to ensure structural stability. After the relaxation of the doped CNT structures a single hydrogen molecule is placed near the impurity and the system is allowed relaxation again. At this point, two *screening criteria* are defined that have to be fulfilled simultaneously: i) the binding energy between the structure and the doping (C-D) must be above 0.6 eV, and ii) the binding energy doping-hydrogen (D-$H_2$) must be between 0.15 and 0.6 eV. To further discriminate between the elements that comply with the criteria, a first indication of *doping performance* is analyzed, to do so, four hydrogen molecules are added near the doping and the average binding energy (D-$H_2$) is checked against the previous criterium.

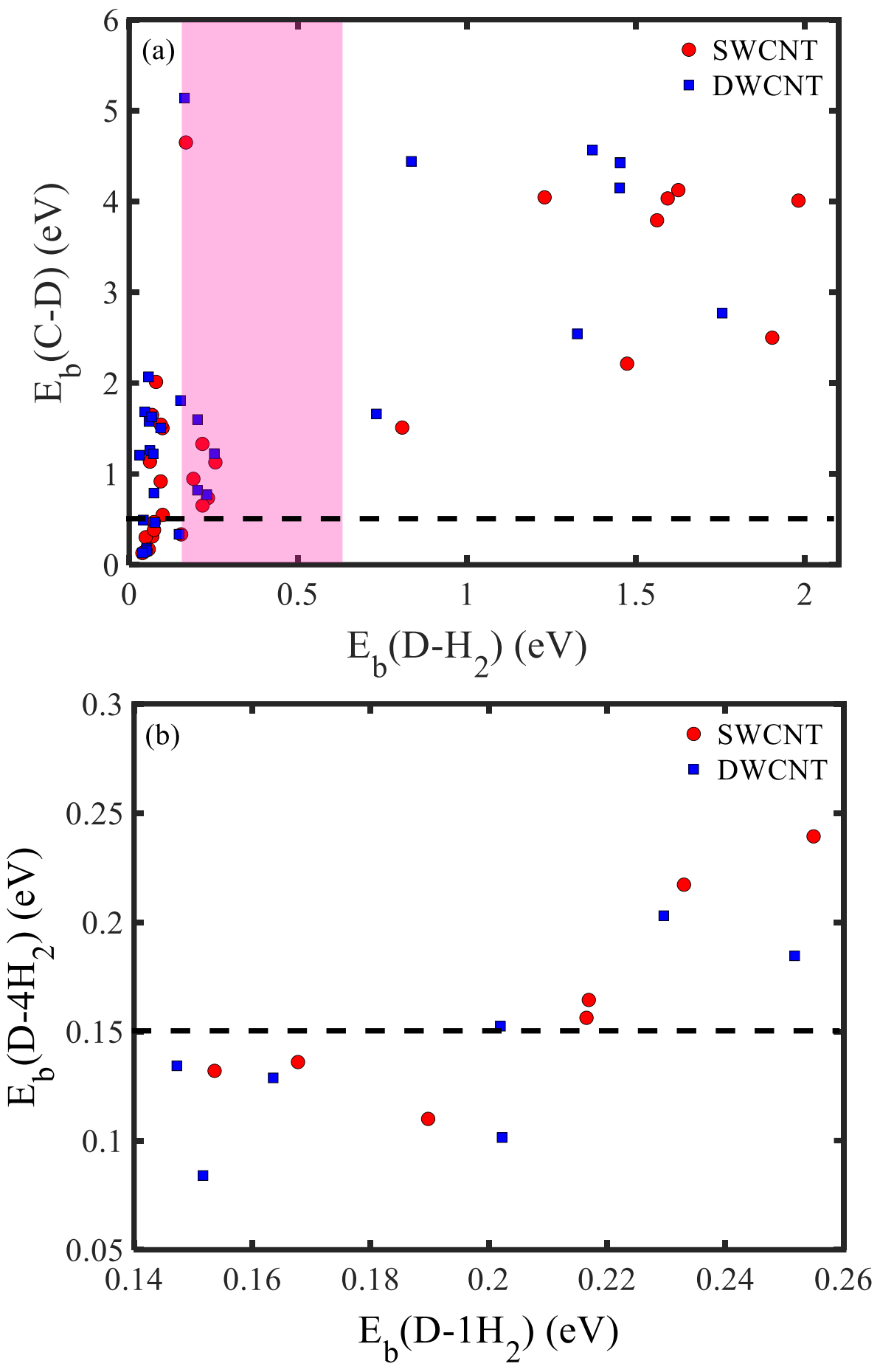

*Figure 3. Binding Energy between (a)impurity and the surface and a hydrogen molecule at the D@CNT surface (b) one and four hydrogen molecules on the surface of doped single and double-walled CNT. SWCNT corresponded to CNT$_{(13,0)}$.*

Figure 3(a) shows the screening criteria of the doped CNTs, the binding energy between the impurities and the external surface of single and double-walled CNTs has been calculated. Binding energies, $E_b$(C-D), above 0.6 eV are considered as acceptable values for doping-carbon interaction (0.6eV<$|E_b|$<1eV strong physical bond, $|E_b|$>1eV chemical bond). The defined range ($|E_b|$>0.6 eV) is especially important when it comes to the long-term applications as it allow cyclability without desorbing the doping from the surface during regeneration, which would lower the performance over time . In Figure 3(a), the dashed-line represents the lower limit (0.6 eV) of binding energy between different impurities and the surface of CNTs and the highlighted region denotes the acceptable range for hydrogen binding energy, $E_b$(D-$H_2$), as defined by the DOE (ranging from 0.15 to 0.6 eV). The elements displayed in the intersection between the regions, fullfilling the selection criteria, represents the promising candidates. The results indicated that the structures doped with Cu, Zn, Au, N, O, F, and Cl were capable of adsorbing hydrogen within the DOE's specified range. Notably, the zinc impurity exhibited a very weak physical bond with the CNT surfaces, rendering it unsuitable for hydrogen storage applications.

The first performance indicator is analyzed to further discriminate the doping. As mentioned before, four hydrogen molecules were added to the surface of CNTs doped with Cu, Zn, Au, N, O, F, and Cl impurities. Figure 3(b) compares the hydrogen-doping binding energy when one and four hydrogen molecules are adsorbed. Some of the elements selected as promising in the previous step are discarded as they do not comply with the binding energy criteria in these conditions. In the cases of Zn, F, O, N, and Cl impurities, the average binding energy between hydrogen molecules and the surface of doped CNTs was either below or at the lower boundary of the DOE range. Consequently, these doped structures were not able to adsorb more than four hydrogen molecules per impurity. Comprehensive data of binding energies can be found in Table S2 of the SI.

Among the impurities under study, the most remarkable results were achieved with the addition of copper and gold atoms to the CNT surfaces. The proposed approach confirmed our previous results regarding Au-CNTs as promising material for reversible hydrogen storage in mild conditions [40]. Therefore, the next sections focus on an extensive investigation on Cu-doped CNTs and its performance as a promising solution to the hydrogen storage challenge.

This structure has demonstrated remarkable effectiveness in various aspects of hydrogen storage. It is worth noting that experimentally, the creation of multi-walled CNTs and metal clusters is virtually possible. Therefore, we examined both of these aspects to extract the necessary information. In the initial step, copper was strategically placed on the surface of CNT in three distinct positions. Following the relaxation of these structures, the results indicated that copper exhibited nearly identical binding energies in all three positions. The most stable configuration was identified when the carbon impurity was positioned at the center of the carbon hexagon. Therefore, this position is selected for the next steps of the simulations. Tested positions and associated total energy can be found in in Figure S1 of the SI.

As shown in Figure 4(a), (b), the partial density of states (PDOS) has been calculated for copper impurity before and after the adsorption onto the CNT surfaces. Upon the addition of copper to the CNT surfaces, substantial alterations were observed in the s and d orbitals. The most notable change was the significant reduction in the number of available states in both the s and d orbitals after copper adsorption. This phenomenon is expected due to the chemisorption of copper onto the CNT surfaces, wherein several empty states in the s and d orbitals of copper became occupied by interacting with carbon orbitals. In the case of d orbitals, a shift towards lower energies was evident, indicating a charge transfer from the CNT surfaces to the copper impurity, resulting in increased stability. These unoccupied states play a crucial role in interacting with hydrogen molecules and facilitating their adsorption. With the introduction of more copper impurities, the number of available states increases, enhancing the potential of this structure for effectively adsorbing hydrogen molecules. Similar behavior, characterized by a diminution in available states and a leftward shift attributable to charge transfer, were observed in Au-doped CNT [40]. Notably, the number of unoccupied states surpassed that observed in copper, a discrepancy anticipated as a consequence of the physical adsorption of Au impurities.

The density of states (DOS) has been calculated for CNTs before and after adding copper impurity (Figure 4(c)). The DOS results for $CNT_{(19,0)}$ and DWCNT are provided in Figure S2 of the SI. As shown before, the addition of copper impurity to the surface of the CNTs results in an increased number of available states across a range of energies, as exemplified by the region around -6 eV in the highlighted section of the DOS (Figure 4(c)). Notably, as the number of copper impurities increases, so does the number of available states as indicated in the highlighted region in Figure 4(c) comparing 1Cu and 2Cu. These additional states enhance the likelihood of these structures participating in physical and chemical reactions. Moreover, a shift to higher energies for the Fermi energy ($E_F$) has been observed after the addition of copper to the surface of CNTs. This shift indicates a charge transfer occurring between copper and the CNT surfaces. Consequently, the CNT structures transition to higher energies, resulting in reduced stability compared to pure CNTs. It's important to note that all copper-doped CNTs remain stable; the term "instability" is used here to compare these structures with pure CNTs. Notably, the $E_F$ shift is higher for DWCNT compared to SWCNTs (Figure S2 in the SI). This observation aligns with the binding energy results (as explained above), Where the binding energy between copper and the surface of DWCNT is stronger than for the SWCNTs ($CNT_{(13,0)}$, $CNT_{(19,0)}$). Therefore, it is expected that the charge transfer between copper and the surface of CNTs is more substantial in DWCNT. It is worth mentioning that instability can be a reason for a structure to be involved in physical and chemical reactions. In these simulations, instability has become a motivation for Cu-doped carbon nanotubes to absorb hydrogen molecules and achieve stability.

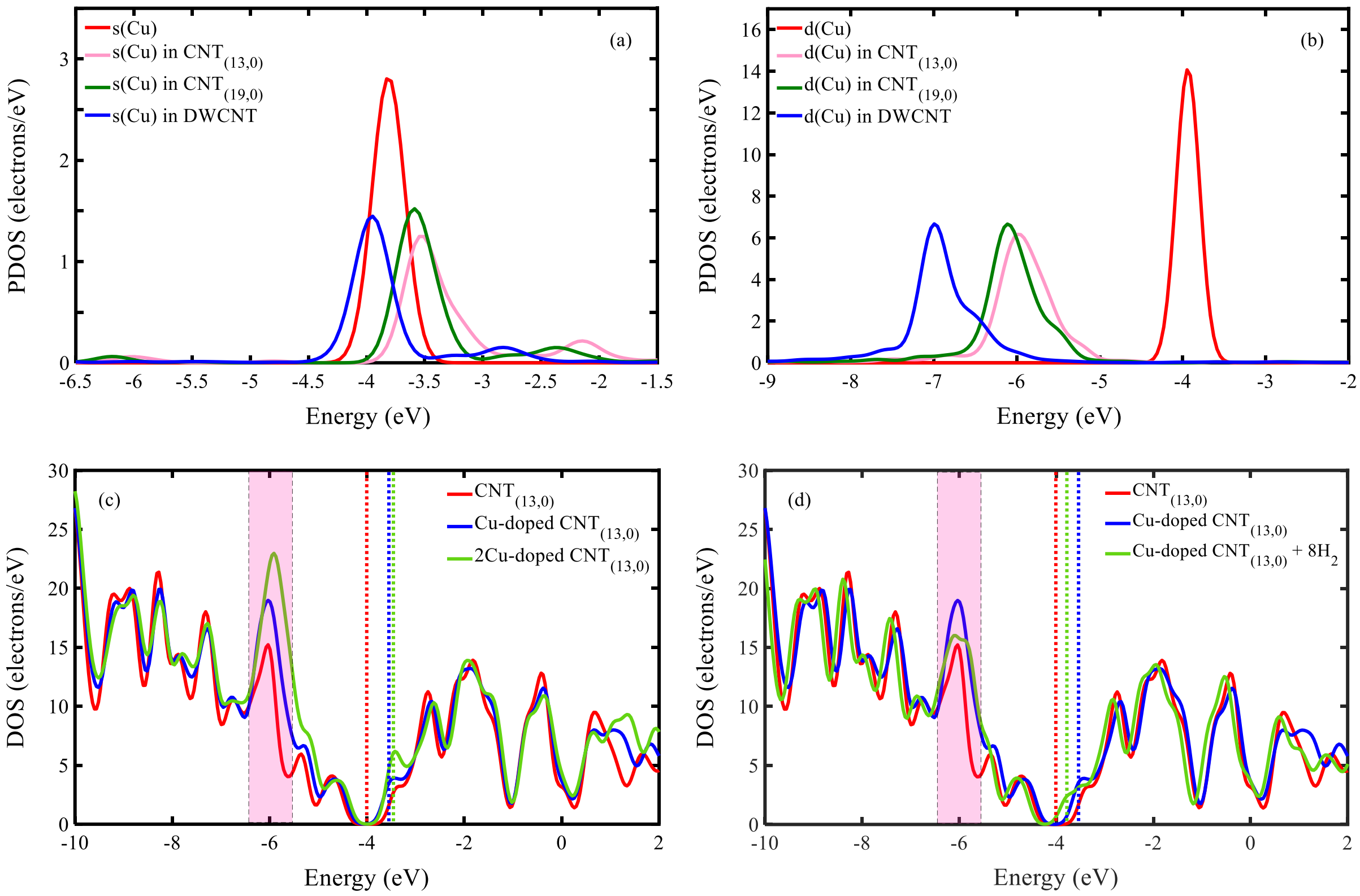

*Figure 4. PDOS of (a) s and (b) d-orbitals of the isolated Cu atom and Cu atom on the surface of single and double-walled CNTs. DOS of the (c) pure and doped $CNT_{(13,0)}$ with one and two copper atoms, (d) Cu-doped, and Cu-doped $CNT_{(13,0)}$ after adsorbing eight hydrogen molecules. Dashed lines indicate the Fermi Energy ($E_F$) per configuration.*

Figure 4(d) displays the density of states (DOS) results for Cu-doped $CNT_{(13,0)}$ after adsorbing eight hydrogen molecules. As depicted in Figure 4(d), the number of available states decreases upon the adsorption of eight hydrogen molecules, as evident around the energy level of -6 eV. In essence, several of the previously empty states are occupied as they interact with hydrogen molecules. Furthermore, a noticeable shift to the left is observed in the Fermi energy, indicating a charge transfer between hydrogen molecules and the surface of $CNT_{(13,0)}$.The charge transfer from the hydrogen molecules to the surface of the $CNT_{(13,0)}$ is reflected in the decrease of total energy of the structure (Cu-doped $CNT_{(13,0)}$ with eight hydrogen molecules). As a result, the structure moves towards stability after the hydrogen adsorption. The same behavior is observed for $CNT_{(19,0)}$ and DWCNT (Figure S3 of the SI). In the case of DWCNT, the Fermi energy shift is almost identical to the SWCNTs, but for adsorbing seven hydrogen molecules, which aligns with the binding energy results. The binding energy between hydrogen and the surface of the DWCNT is weaker compared to SWCNTs. This behavior can be attributed to the binding energy between copper and the CNT surfaces, where binding energies of 1.22 eV, 1.127 eV, and 1.121 eV have been calculated for DWCNT, $CNT_{(13,0)}$, and $CNT_{(19,0)}$, respectively. With stronger binding energy between copper and the CNT surface, the number of available states decreases, leading to reduced interaction with hydrogen. This relationship is further elucidated by the total energy per atom for each structure, where a more stable structure exhibits less motivation to engage in physical and chemical reactions. The total energy per atom for Cu-doped DWCNT is lower at -172.34 eV/atom compared to Cu-doped SWCNTs ($CNT_{(13,0)}$ = -169.21 eV/atom and $CNT_{(19,0)}$ = -166.36 eV/atom). As a result, Cu-doped DWCNT is more stable than SWCNTs, hence demonstrating less motivation to be involved in physical and chemical reactions, which results in weaker hydrogen adsorption compared to SWCNTs. This behavior is consistent with the number of adsorbed hydrogen molecules per copper atom (Figure 5), with Cu-doped SWCNTs adsorbing about eight hydrogens, while this number reduces to seven for DWCNT. This reduction is due to the lower motivation of DWCNT to engage in physical and chemical reactions, a fact corroborated by the DOS results after hydrogen molecule adsorption. As indicated in the DOS results, the shift in Fermi energy for DWCNT after adsorbing seven hydrogen molecules is nearly equivalent to the Fermi energy shift for SWCNTs after adsorbing eight hydrogen molecules (Figure 4(d) and Figure S3 of the SI). This suggests that the charge transfer and binding energy are weaker between hydrogen molecules and the DWCNT surface. Additionally, for $CNT_{(19,0)}$, the binding energy and charge transfer are more pronounced compared to $CNT_{(13,0)}$. Thus, increasing the diameter of the CNT has a positive impact on hydrogen adsorption. It is anticipated that for larger CNT sizes, the number of adsorbed hydrogen molecules will increase, this conclusion aligns well with the results obtained for Au-CNTs [40].

An essential consideration to bear in mind is that, even after adsorbing eight and seven hydrogen molecules onto the surfaces of single and double-walled CNTs, the binding energy between copper and the CNT surfaces remains a robust chemical bond. Binding energies of approximately 1.20 eV and 1.10 eV have been calculated for DWCNT and SWCNTs, respectively. This is a pivotal aspect for a structure intended for numerous hydrogen adsorption-desorption cycles. In practical terms, the structure of Cu-doped CNTs remains unaltered even after adsorbing hydrogen molecules. The chemical bond between copper and the CNT surfaces strikes a balance -neither excessively strong to alter the properties of the CNTs nor overly weak to result in structural collapse after multiple adsorption- desorption cycles.

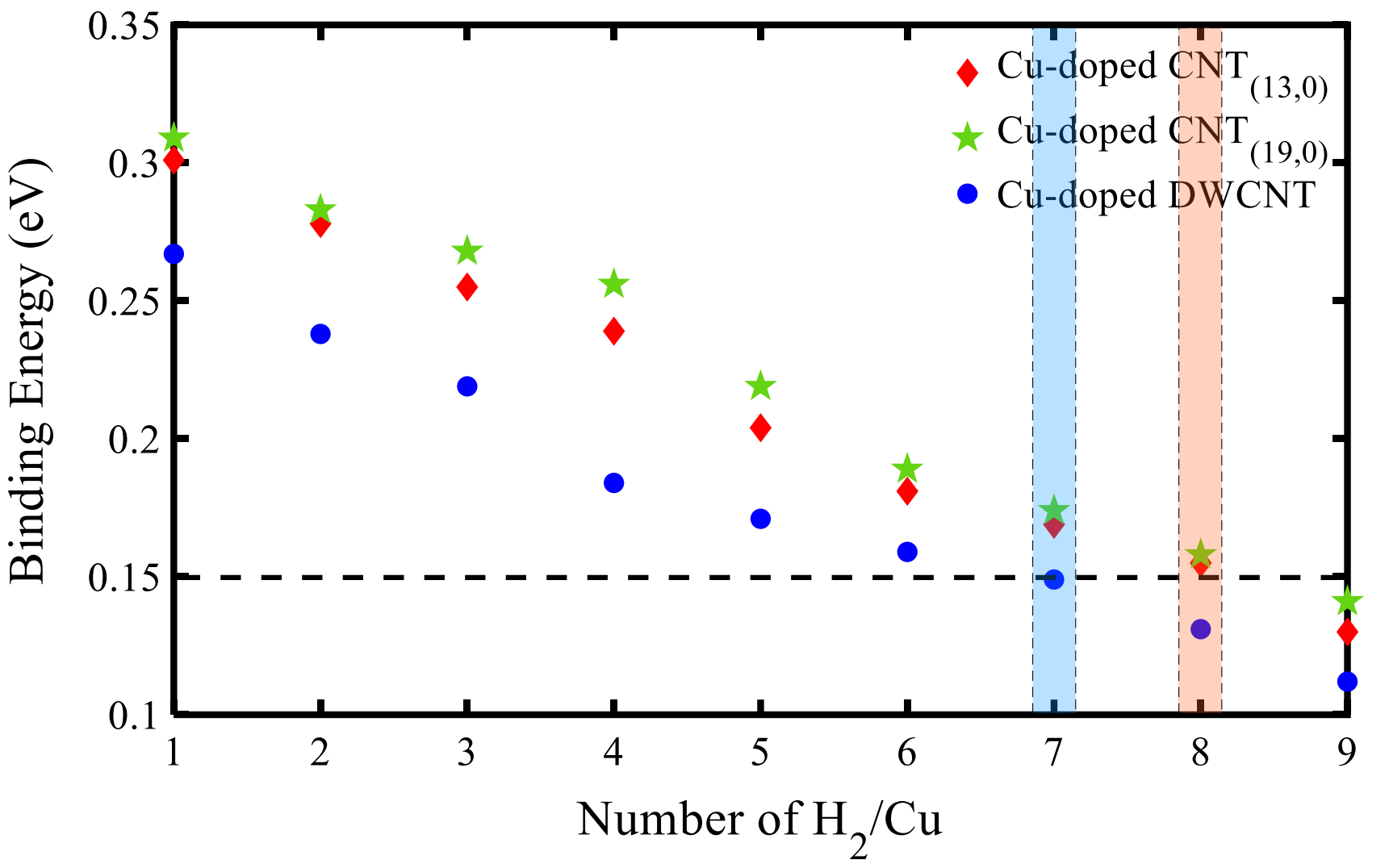

*Figure 5. Binding energy between hydrogen molecules and the surface of CNTs.*

Charge transfer has been calculated by Eq. (3) to interpret the adsorption of copper atoms and molecular hydrogen on the surface of CNTs. The charge transfer results reveal values of -0.08 e, -0.07 e, and -0.18 e for $CNT_{(13,0)}$, $CNT_{(19,0)}$, and DWCNT, respectively. The negative values indicate charge transfer occurring from the surfaces of both single and double-walled CNTs to the copper impurity. This finding is consistent with the PDOS and DOS results. The observed displacement of the Fermi energy suggested that charge transfer from the CNT surfaces to the copper impurity had indeed transpired. The obtained values of charge transfer indicate that the interaction of copper intensifies with an increase in the diameter and the number of carbon layers in CNTs. Consequently, it is anticipated that the binding energy between copper and the CNT surfaces will be considerably stronger in the case of DWCNT. However, as the interaction of copper with the CNT surface intensifies, the number of available states in the copper impurity diminishes. Many of these available states of copper become occupied through interactions with carbon atoms. Therefore, the number of available states in Cu-doped CNTs decreases, resulting in reduced hydrogen molecule adsorption, especially remarkable in the case of DWCNT, where the value of charge transfer is pronounced.

The results are in alignment with the binding energy outcomes. As depicted in Figure 5, Cu-doped DWCNT exhibits a reduced number of adsorbed hydrogen molecules, reflecting the fewer available states in the copper impurity. Hence, although the binding energy between the copper impurity and the DWCNT surface (1.22 eV) surpasses that between the copper impurity and the SWCNT surfaces (approximately 1.12 eV), the number of adsorbed hydrogen molecules on DWCNT diminishes. It's important to note that the binding energy between copper and the surfaces of single and double-walled CNTs is classified as chemisorption. This implies that both are well-suited for long-term applications and having a stronger chemical bond does not render DWCNTs superior to SWCNTs. The charge transfer resulting from the adsorption of a single hydrogen molecule on Cu-doped CNTs is approximately 0.03 e for both $CNT_{(13,0)}$ and $CNT_{(19,0)}$, and 0.02 e for DWCNT. This finding aligns with the binding energy results following the adsorption of a single hydrogen molecule on Cu-doped CNTs (as shown in Figure 5). After the adsorption of eight hydrogen molecules on $CNT_{(13,0)}$ and $CNT_{(19,0)}$, the charge transfer has increased to 0.15 e. A nearly equivalent amount of charge transfer, 0.149 e, is observed following the adsorption of seven hydrogen molecules on DWCNT. Notably, these values are positive, indicating that the charge transfer occurs from hydrogen molecules to the surfaces of both single and double-walled CNTs. This is in concordance with the $E_F$ displacement depicted in Figure 4(d).

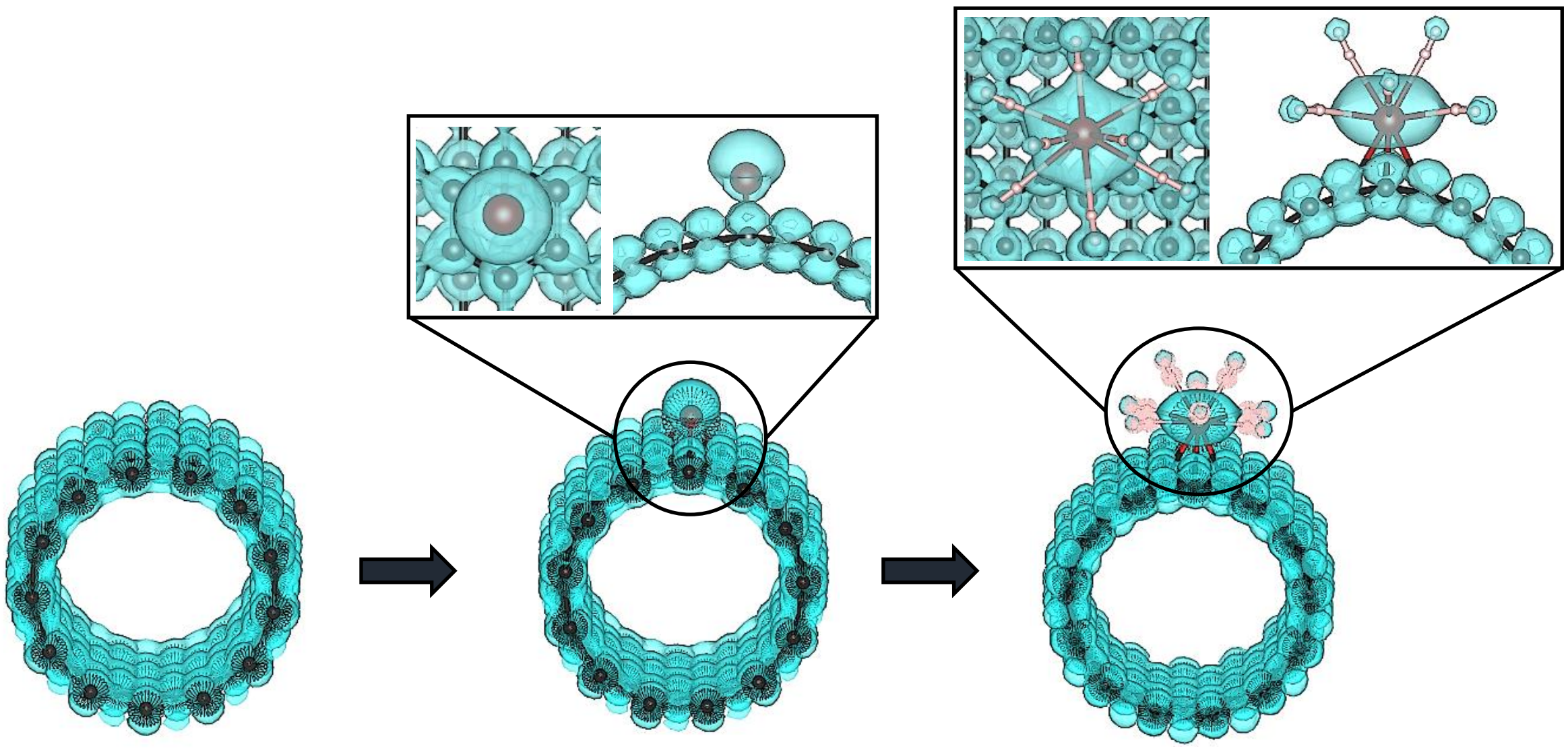

*Figure 6. LDOS of pure, Cu-doped, and Cu-doped $CNT_{(13,0)}$ after adsorbing eight hydrogen molecules.*

Figure 6 illustrates the distribution of charge in pure CNT, Cu-doped CNT, and Cu-doped CNT after the adsorption of eight hydrogen molecules. In pure CNT, charge is uniformly distributed, a testament to its stability. The addition of copper doping to the CNT surface causes a slight alteration in the distribution of charge around the copper impurity, an expected change due to the chemical bond formed between copper and the CNT surface. Remarkably, this chemical bond does not compromise the stability of the CNT structure, and the new Cu-doped CNT structure exhibits uniform charge distribution. However, after the adsorption of eight hydrogen molecules on the Cu-doped CNT surface, the charge distribution undergoes significant changes, particularly around the copper impurity. A substantial portion of the charge from hydrogen molecules is transferred to the copper impurity, in line with the charge transfer results. As per the charge transfer findings, this charge transfer occurs from hydrogen molecules to the CNT surface. Moreover, the entire Cu-doped CNT structure is expected to stabilize due to the acquisition of charges from hydrogen molecules. This behavior is evident in the $E_F$ displacement within the DOS diagram (Figure 4(c), (d)). The distribution of charge across the entire CNT structure remains uniform, further attesting to the stability of this structure throughout the impurity and hydrogen adsorption process.

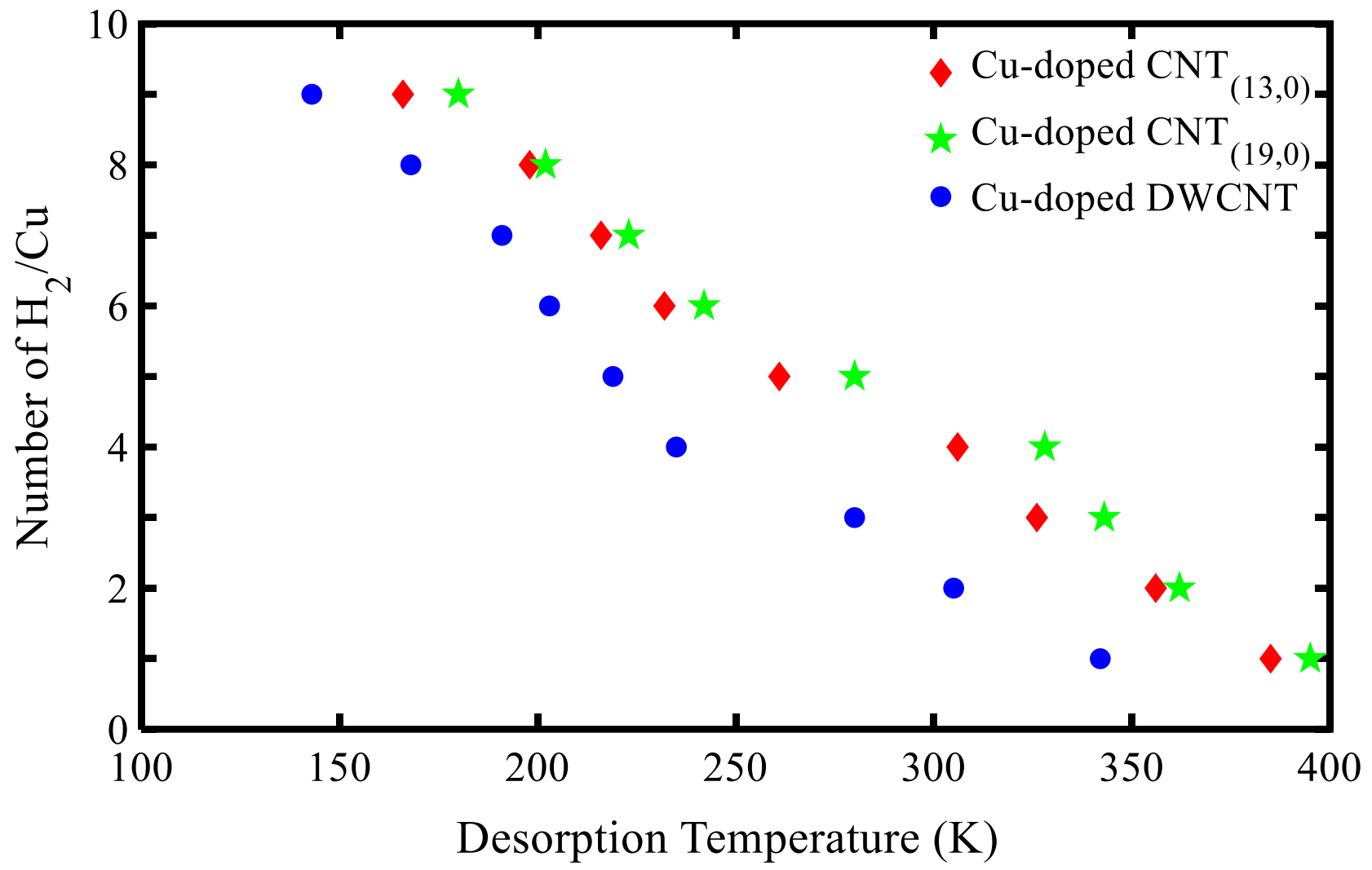

*Figure 7. Desorption Temperature of Cu-doped CNT$_{(13,0)}$, CNT$_{(19,0)}$, and DWCNT.*

Desorption temperatures were computed to assess the release of 1 to 9 hydrogen molecules per impurity from the surfaces of Cu-doped CNT$_{(13,0)}$, CNT$_{(19,0)}$, and DWCNT using Eq. (4) (see Figure 7). As depicted in Figure 7, an increase in the number of hydrogen molecules corresponds to a decrease in desorption temperature at isobaric conditions of 1 bar. This relationship can be attributed to the direct dependence of desorption temperature on adsorption energy. In simpler terms, as the number of hydrogen molecules increases, adsorption energy decreases, leading to a lower desorption temperature. Temperature lifts of about 200 K are estimated for the total regenration of the structures at isobaric conditions, stablishing the operational range of Cu-doped CNTs. Desorption temperatures of 198 K, 200 K, and 191 K were calculated for initiating the release when assuming maximum hydrogen capacity per copper atom (eight and seven hydrogen molecules for SWCNTs and DWCNT, respectively) from the surfaces of CNT$_{(13,0)}$, CNT$_{(19,0)}$, and DWCNT, respectively (Figure 7). The desorption temperature results suggest that Cu-doped CNTs can serve as suitable structures for reversible hydrogen storage applications at moderate temperatures (non-cryogenic) and ambient pressure.

The formation of metal clusters on the surface of CNTs is a common phenomenon during the doping addition process. Therefore, it is valuable to assess the performance of CNTs after the formation of copper clusters. Figure S4 of the SI illustrates the addition of copper clusters of four different sizes, containing three, six, nine, and twelve copper atoms ($Cu_3$, $Cu_6$, $Cu_9$, and $Cu_{12}$), to the surfaces of $CNT_{(13,0)}$. The results for $CNT_{(19,0)}$ and DWCNT can be found in Figure S5 and Figure S6 of the SI and a compilation of the binding energies in Table S3. These clusters were initially relaxed and subsequently added to the CNT surfaces. Hydrogen molecules were then introduced to evaluate the performance of these new structures for hydrogen adsorption. $CNT_{(13,0)}$, $CNT_{(19,0)}$, and DWCNT were exposed to fourteen, twenty-two, twenty-four, and thirty hydrogen molecules, for cluster size of $Cu_3$, $Cu_6$, $Cu_9$, and $Cu_{12}$, respectively.

In the case of $Cu_3$-$CNT_{(13,0)}$, ten out of fourteen hydrogen molecules were adsorbed with an average binding energy of about 0.19 eV. A similar number of hydrogen molecules, with a comparable binding energy, were adsorbed onto the surface of $CNT_{(19,0)}$. This number decreased to nine hydrogen molecules with an average binding energy of approximately 0.21 eV for DWCNT. Consequently, SWCNTs demonstrated superior performance in comparison to DWCNT. Additionally, the binding energy between copper clusters and the CNT surfaces was calculated to assess the suitability of these structures for long-term applications.

Based on the results, after adsorbing ten and nine hydrogen molecules onto the surfaces of single and double-walled $Cu_3$-CNTs, average binding energies of 1.86 eV, 1.78 eV, and 2.06 eV were calculated between the copper cluster ($Cu_3$) and the surfaces of $CNT_{(13,0)}$, $CNT_{(19,0)}$, and DWCNT, respectively. This difference in binding energy is likely due to the stronger bond between the copper cluster and the surface of DWCNT, which is expected to result in a reduced number of available states in DWCNT compared to SWCNTs.

It is expected that the number of available states for adsorbing hydrogen molecules in DWCNT is less than in SWCNTs due to stronger bonding between the copper cluster and the surface of DWCNT. In all cases, the bond between the copper clusters and the CNT surfaces remained a strong chemical bond, affirming the potential of these structures for long-term applications.

Increasing the cluster size, $Cu_6$-CNT resulted in the adsorption of eighteen hydrogen molecules on the surface of $CNT_{(13,0)}$ with an average binding energy of approximately 0.19 eV. Similar results were observed for $CNT_{(19,0)}$. In the case of DWCNT, sixteen hydrogen molecules with an average binding energy of about 0.24 eV were adsorbed. The binding energies between the $Cu_6$ cluster and the surfaces of $CNT_{(13,0)}$, $CNT_{(19,0)}$, and DWCNT were calculated as 1.41 eV, 1.35 eV, and 1.69 eV, respectively. Notably, the binding energy for DWCNT was stronger, indicating fewer available empty orbitals and, consequently, less hydrogen adsorption.

In the case of $Cu_9$-CNT, sixteen hydrogen molecules were adsorbed to the surface of $CNT_{(13,0)}$, with an average binding energy of about 0.57 eV. The binding energy between $Cu_9$ cluster and the surface of $CNT_{(13,0)}$ reached 1.64 eV. In $CNT_{(19,0)}$, the number of adsorbed hydrogen molecules reached nineteen, with an average binding energy of about 0.26 eV (Figure S5(c) of the SI). The binding energy for $CNT_{(19,0)}$ was calculated as 1.23 eV. These results highlight the positive impact of increasing the diameter of CNTs on hydrogen adsorption in agreement with the resuts obtained with Au-doped CNTs [40]. Despite the addition of similar initial configuration of copper clusters to the CNT surfaces, the doping-CNTs interaction is diamter dependent, affecting the most stable configuration of the clusters on the CNTs, shape, and accessible surface area. For $CNT_{(19,0)}$, the shape of the copper cluster provided more accessibel adsorption sites compared to the other CNTs under study, leading to increased capacity of hydrogen. In the case of DWCNT, sixteen hydrogen molecules were adsorbed with an average binding energy of 0.50 eV. The binding energy between the copper cluster and the surface of DWCNT was calculated as 1.9 eV.

Further increasing the cluster size, $Cu_{12}$-CNT resulted in the adsorption of twenty-two hydrogen molecules, with an average binding energy of 0.37 eV, on the surface of $CNT_{(13,0)}$. Additionally, a binding energy of 1.07 eV was calculated between the copper cluster and the surface of $CNT_{(13,0)}$. In the case of $CNT_{(19,0)}$, the number of adsorbed hydrogen molecules increased to twenty-four, with an average binding energy of 0.48 eV between hydrogen molecules and the surface of $CNT_{(19,0)}$ and binding energy cluster-CNTs of 1.06 eV for $CNT_{(13,0)}$ and $CNT_{(19,0)}$. The differing adsorption of hydrogen molecules can be attributed to the shape of the cluster. As shown in Figure S5 of the SI, the $Cu_{12}$ cluster attached to the surface of $CNT_{(19,0)}$ involved two copper atoms, while three are observed in the case of $CNT_{(13,0)}$ (Figure S4). Hence, the difference in the configuation of the cluster provided more accessible states and surface area for hydrogen adsorption in $Cu_{12}$-$CNT_{(19,0)}$. For DWCNT, twenty-two hydrogen molecules were adsorbed with an average binding energy of about 0.45 eV. Similar to $CNT_{(19,0)}$, two copper atoms were involved in attaching the copper cluster to the surface of DWCNT, while the binding energy between the copper cluster and the DWCNT surface remained a strong chemical bond, with an average binding energy of 1.77 eV.

Figure 8 illustrates the average hydrogen binding energy concerning the cluster size and its correlation with the desorption temperature required to initiate hydrogen recovery at ambient pressure. As anticipated, the cluster size influences the material's performance within operational condition ranges. The increase in cluster size exhibits a non-linear behavior with the binding energy of the adsorbed hydrogen but demonstrates a clear relation with the Cu/H2 ratio. With the growth in cluster size, the amount of hydrogen adsorbed per Cu atom decreases, as the available states are now occupied by Cu-Cu interactions. Simultaneously, the desorption temperature is contingent on the binding energy (Eq. (4)), emphasizing that controlling the cluster size could enable the fine-tuning of operational conditions. However, a trade-off between high working temperatures and gravimetric capacity is evident from the Cu/$H_2$ ratio. Figure 8 highlights that CNTs doped with Cu1-Cu6 exhibit the best performance concerning capacity and near-ambient working conditions.

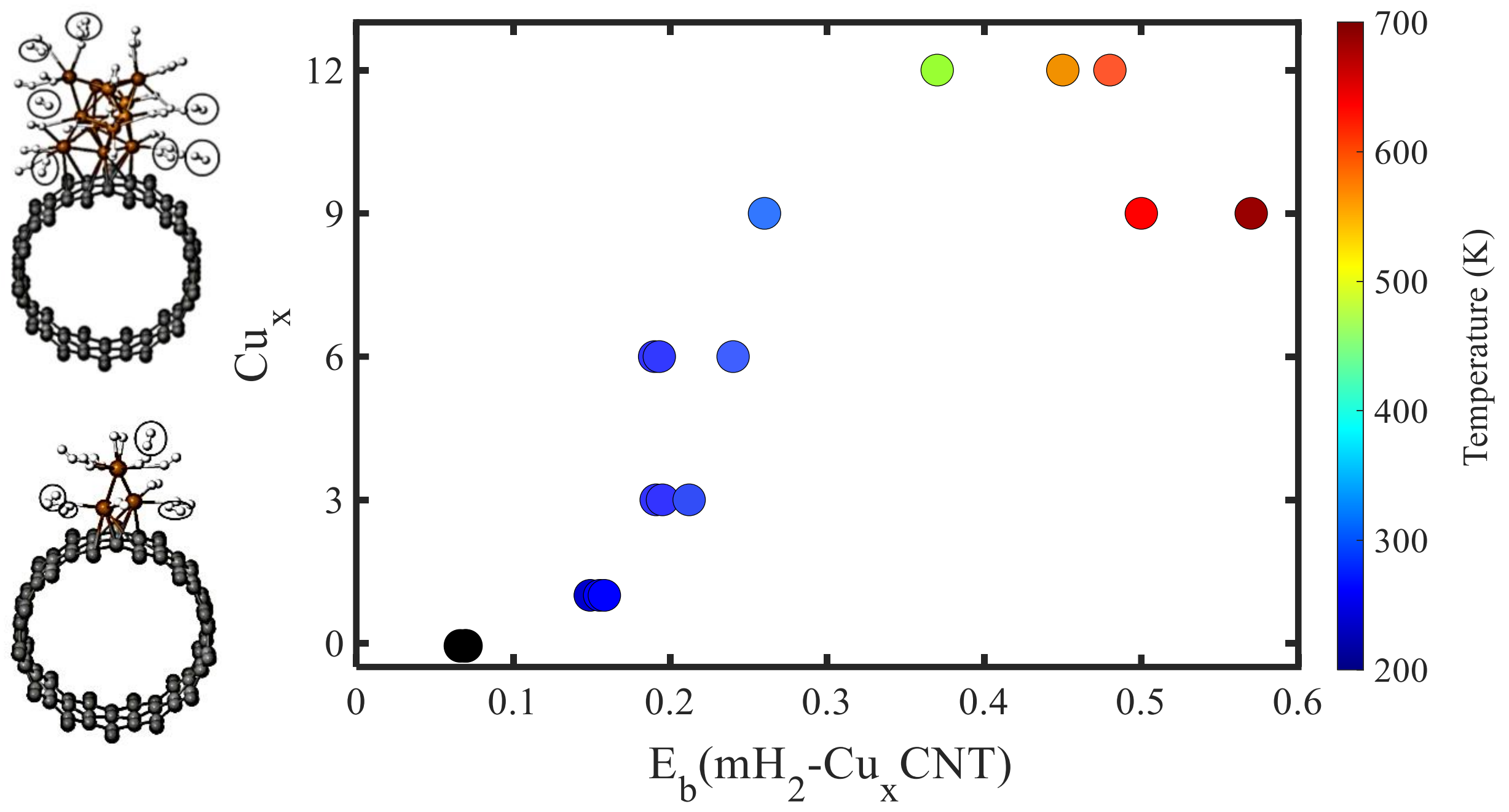

*Figure 8. Average binding energy (D-H2) for maximum adsorbed hydrogen as a function of the cluster size. The color scale indicates the desorption temperature at ambient pressure.*

In the final step of the analysis, the gravimetric capacity was computed using Eq. (5), which stands as a crucial parameter for assessing the suitability of materials for hydrogen storage applications. In the case of $CNT_{(13,0)}$ and $CNT_{(19,0)}$, it was observed that up to eight hydrogen molecules per copper atom could be adsorbed. The outcomes of the gravimetric capacity assessment unveiled that approximately six copper atoms on the surface of $CNT_{(13,0)}$ were required to achieve a gravimetric capacity of approximately 5.60 wt%. Consequently, introducing approximately 6% copper impurities to $CNT_{(13,0)}$ could facilitate the attainment of the target set by the DOE. In the case of $CNT_{(19,0)}$, approximately eight copper atoms were found to yield a gravimetric capacity of approximately 5.23 wt%. This implies that introducing approximately 5% copper impurity is sufficient to meet the DOE defined target. In order to substantiate the assessment, a DFT simulation was conducted on $CNT_{(13,0)}$. Within this simulation, approximately 60 hydrogen molecules were introduced onto the surface of the CNT structure. Following a relaxation process, it was observed that approximately 48 of these hydrogen molecules became adsorbed by copper impurities (8 hydrogen molecules per copper). The corresponding results are presented in Figure S7 of the Supplementary Information (SI). It is noteworthy that the arrangement or configuration of copper impurities can be change. The primary objective should be achieving a copper impurity content of approximately 6%.

However, when considering DWCNTs with a hydrogen adsorption ratio of seven molecules per copper atom, it was discerned that around fifteen copper atoms were necessary to attain a gravimetric capacity of approximately 5 wt%. This, in turn, corresponds to an approximate 6% copper impurity level. However, it is difficult to achieve this configuration of copper atoms in DWCNTs due to the comparable surface area of DWCNTs to $CNT_{(19,0)}$. The risk of aggregation of copper atoms further complicate the matter and challenges the understanding of the required distribution of copper atoms. As a result, it can be concluded that SWCNTs are better than DWCNTs in terms of hydrogen storage performance.

It is worth noting that the functionalization of CNTs has been performed by concentrated on their outer layers. Recent technological advancements have enabled researchers to introduce impurities within the interior of CNTs [47]–[49]. The introduction of such impurities within CNTs has the potential to significantly increase the gravimetric capacity of these structures, surpassing the values assessed in the present study.

To assess the effectiveness of the proposed approach, particularly in discriminating impurities based on binding energy, two key elements are analyzed: i) the desorption temperature of a single hydrogen on the D@CNTs exhibiting binding energies above the DOE range, and ii) the gravimetric capacity of the D@CNTs preselected in the screening criteria but discarded after the preliminary performance test.

Figure S8 depicts the binding energy (C-D) as a function of the desorption temperature of hydrogen for systems exhibiting strong interaction (above 0.6 eV). It is evident that high temperatures are required to achieve complete regeneration. all elements in this behavior group show desorption temperatures above 900 K. While these desorption temperatures might be desirable, such working conditions would compromise the stability of the CNTs, as reported in previous studies around 1000 K in the air [50], making them unsuitable for the application. On the other hand, the gravimetric capacity of the elements that fulfilled the screening criteria but were disregarded is calculated. For this assessment, CNT(13,0) doped with 6% impurities is used to calculate the gravimetric capacity based on the maximum amount of hydrogen it can absorb within the DOE range. The hierarchy in the number of molecular hydrogens adsorbed per doping is $N(1H_2) < O(3H_2) < Zn=F=Cl\ (4H_2) < Cu\ (8H_2) < Au(9H_2)$. The calculated gravimetric capacity for these conditions results in 0.15 wt% (N), 2.62 wt% (O), 2.8 wt% (Zn), 3.2 wt% (Cl), 3.42 wt% (F), 4.28 wt% (Au), and 5.6 wt% (Cu). These results emphasize that the initial performance assessment was a necessary step to ensure reaching the DOE target. Au and Cu atoms emerged as candidates from the approach, with their gravimetric capacities serving as a clear indication of their performance, with Au-doped CNT slightly below the target under the chosen conditions, even when the ratio of adsorbed molecules per impurity is higher. This further justifies the preselected dopants with lower molecular weight in the initial step of the approach.

Therefore, the proposed strategy can offer reliable information when deciding on the dopants for porous materials, whether for computational or experimental design, providing exploratory tools for making informed decisions based on preliminary performance for hydrogen storage.

## Conclusion

This study introduces a systematic approach for the rational selection of optimal doping on porous materials, emphasizing the interplay between doping, material structure, and hydrogen interaction. The proposed approach involves a step-by-step process: (i) preselection of lightweight dopants, (ii) initial dopant screening based on the stability of the doped structures, (iii) analysis of binding energies in doping-structure and hydrogen-doping interactions, and (iv) a preliminary performance test. As a case study, the strategy was tested in CNTs. The outcome of applying the approach highlights Cu and Au as promising candidates for hydrogen storage.

In-depth analysis for Cu-doped CNTs has been performed using key parameters such as PDOS, charge transfer, desorption temperature profiles, and a detailed evaluation of gravimetric capacity. The cluster formation of the added dopings has been considered and analyzed for small configurations ranging from $Cu_3$ to $Cu_{12}$. The effect of the cluster size presents a non-linear behavior with the binding energy of hydrogen and working temperature, with an evident and critical influence on the capacity in terms of the adsorbed amount per Cu atom. Our study suggests a new direction for future research, exploring cluster size control by tuning the distance between CNTs and tailoring performance. The strong chemical bond between the copper impurity and the CNT surface, even after hydrogen adsorption, underscores the stability of these structures for long-term applications. Gravimetric capacity calculations reveal competitive performance, achieving the DOE-defined target with an impurity content ranging from 5% to 6% for SWCNTs. Gravimetric capacity has been calculated under the same conditions for previously disregarded dopings, evidencing Cu-doped CNTs as outperforming candidates among the studied systems.

The evaluation of our proposed strategy demonstrates that this approach serves as a robust framework for assessing and predicting the performance of doped materials, offering valuable insights for both computational and experimental design. As the next steps, testing and assessing the method when doping is placed on the internal surface of the pores is recommended, offering valuable insights for the design of advanced hydrogen storage materials.

## Conflicts of interest

The authors declare no conflicts of interest.

## Acknowledgment

EIRES – Eindhoven Institute for Renewable Energy Systems.

# ***Supplementary Information***

## Table of Contents **Page**

*Table S1. Glossary*

| Symbol | Explanation | units |
|---|---|---|
| $E_b$ | Binding energy | eV |
| $E_F$ | Fermi energy | eV |
| $E_{tot(CNT)}$ | Total energy of carbon nanotube per unit cell | eV |
| $E_{tot(Cu)}$ | Total energy of isolated copper atom | eV |
| $E_{tot(Cu-doped\ CNT)}$ | Total energy of carbon nanotube doped with a copper atom per unit cell | eV |
| $E_{tot(iH_2)}$ | Total energy of *i* hydrogen molecules | eV |
| $E_{tot((Cu-doped\ CNT)+iH_2)}$ | Total energy of the i adsorbed hydrogen molecules on the surface of carbon nanotube doped with a copper atom per unit cell | eV |
| $E_{ads}$ | Adsorption energy | J/mol |
| $Q_t$ | Charge transfer | e |
| $Q_{(ads-H_2)}$ | Total charge of the adsorbed hydrogen molecule | e |
| $Q_{(iso-H_2)}$ | Total charge of the isolated hydrogen molecule | e |
| $T_D$ | Desorption temperature | K |
| wt % | Gravimetric storage capacity | - |
| $W_{H_2}$ | Molecular weight of hydrogen molecule | g/mol |
| $W_{(Cu-doped\ CNT)}$ | Molecular weight of Cu-doped CNT | g/mol |
| $i$ | Number of hydrogen molecules | - |
| $k_B$ | Boltzmann constant | $1.38 \times 10^{-23}\ J/K$ |
| R | Gas constant | 8.314 J/(mol K) |
| $p$ | Atmospheric pressure | 1 atm |
| $\Delta S$ | Change in hydrogen entropy from gas to liquid phase | 75.44 J/(mol.K) |

*Table S2. Binding energy (eV) between different impurities and the surface of single and double-walled CNTs. Binding energy (eV) between one and four hydrogen molecules and the surface of doped single and double-walled CNTs.*

| | **Li** | **Na** | **K** | **Be** | **Mg** | **Ca** | **Sc** | **Ti** | **V** | **Cr** |
|---|---|---|---|---|---|---|---|---|---|---|
| **$E_b$(C-D)** | | | | | | | | | | |
| **SWCNT** | 0.548 | 0.311 | 0.302 | 0.170 | 0.152 | 0.129 | 1.510 | 2.499 | 2.213 | 4.044 |
| **DWCNT** | 0.788 | 0.464 | 0.492 | 0.176 | 0.153 | 0.129 | 1.661 | 2.768 | 2.540 | 4.440 |
| **$E_b$(D-$H_2$)** | | | | | | | | | | |
| **1$H_2$+doped SWCNT** | 0.098 | 0.068 | 0.048 | 0.057 | 0.049 | 0.039 | 0.807 | 1.902 | 1.473 | 1.229 |
| **1$H_2$+doped DWCNT** | 0.073 | 0.075 | 0.042 | 0.050 | 0.051 | 0.039 | 0.731 | 1.754 | 1.325 | 0.834 |
| **4$H_2$+doped SWCNT** | - | - | - | - | - | - | - | - | - | - |
| **4$H_2$+doped DWCNT** | - | - | - | - | - | - | - | - | - | - |
| | | | | | | | | | | |
| | **Mn** | **Fe** | **Co** | **Ni** | **Cu** | **Zn** | **Au** | **B** | **Al** | **Si** |
| **$E_b$(C-D)** | | | | | | | | | | |
| **SWCNT** | 4.123 | 4.008 | 4.032 | 3.791 | 1.127 | 0.334 | 0.734 | 1.503 | 1.135 | 2.012 |
| **DWCNT** | 4.564 | 4.428 | 4.424 | 4.148 | 1.223 | 0.335 | 0.769 | 1.506 | 1.206 | 2.069 |
| **$E_b$(D-$H_2$)** | | | | | | | | | | |
| **1$H_2$+doped SWCNT** | 1.624 | 1.980 | 1.593 | 1.561 | 0.254 | 0.153 | 0.232 | 0.098 | 0.061 | 0.079 |
| **1$H_2$+doped DWCNT** | 1.370 | 1.452 | 1.452 | 1.451 | 0.251 | 0.147 | 0.229 | 0.092 | 0.030 | 0.056 |
| **4$H_2$+doped SWCNT** | - | - | - | - | 0.239 | 0.131 | 0.217 | - | - | - |
| **4$H_2$+doped DWCNT** | - | - | - | - | 0.184 | 0.134 | 0.203 | - | - | - |
| | | | | | | | | | | |
| | **Ge** | **N** | **P** | **As** | **O** | **S** | **Se** | **F** | **Cl** | |
| **$E_b$(C-D)** | | | | | | | | | | |
| **SWCNT** | 1.649 | 4.648 | 0.917 | 1.541 | 0.945 | 0.469 | 0.383 | 1.330 | 0.652 | |
| **DWCNT** | 1.683 | 5.136 | 1.257 | 1.577 | 1.808 | 1.628 | 1.221 | 1.598 | 0.820 | |
| **$E_b$(D-$H_2$)** | | | | | | | | | | |
| **1$H_2$+doped SWCNT** | 0.067 | 0.167 | 0.093 | 0.093 | 0.189 | 0.072 | 0.073 | 0.216 | 0.216 | |
| **1$H_2$+doped DWCNT** | 0.046 | 0.163 | 0.061 | 0.058 | 0.151 | 0.066 | 0.070 | 0.202 | 0.201 | |
| **$4H_2$+doped SWCNT** | - | 0.135 | - | - | 0.109 | - | - | 0.156 | 0.164 | |
| **$4H_2$+doped DWCNT** | - | 0.128 | - | - | 0.084 | - | - | 0.101 | 0.152 | |

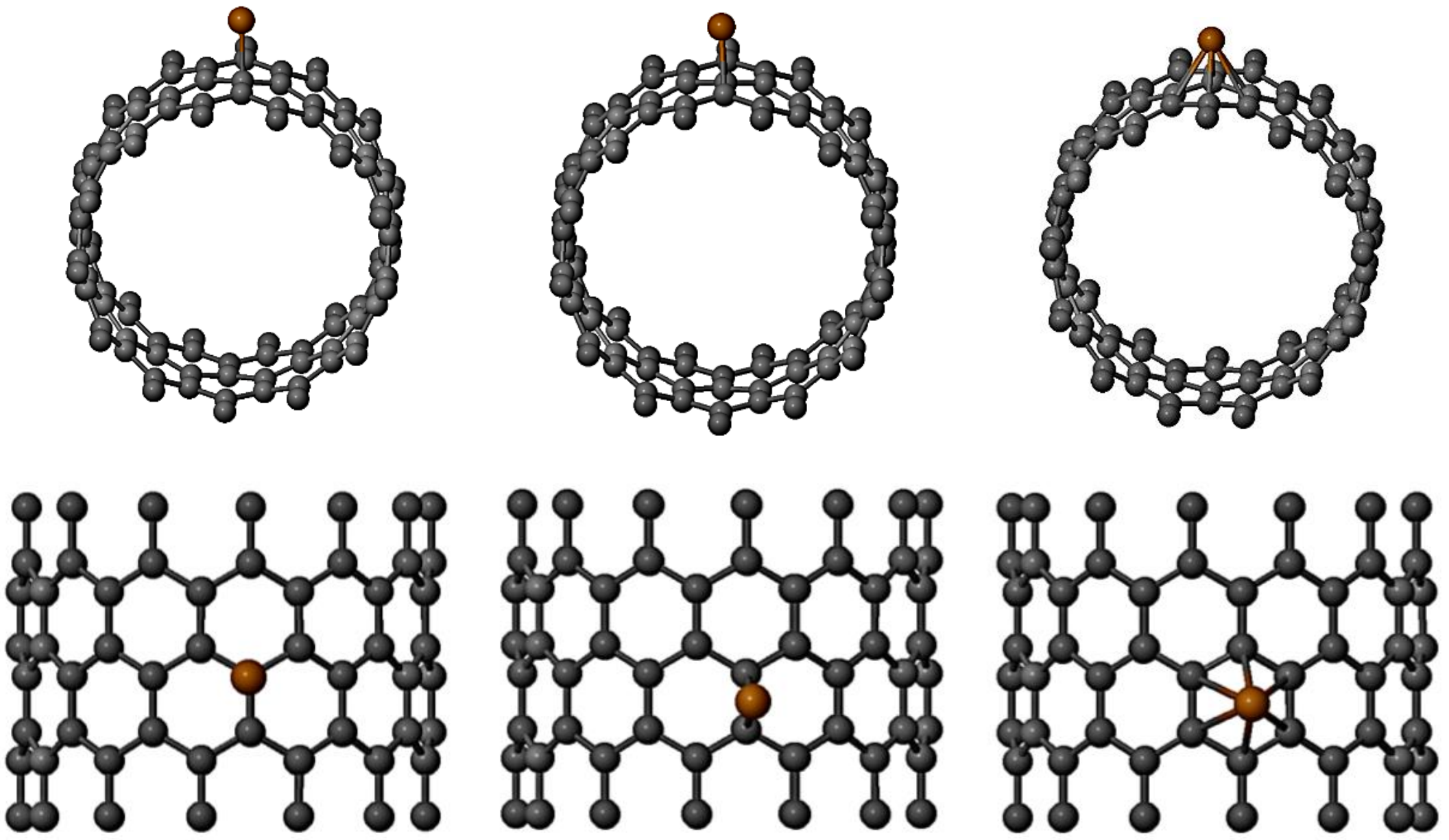

Total energy: -18100.547 eV Total energy: -18100.551 eV Total energy: -18100.557 eV

*Figure S1. Different positions of copper on the surface of carbon nanotube.*

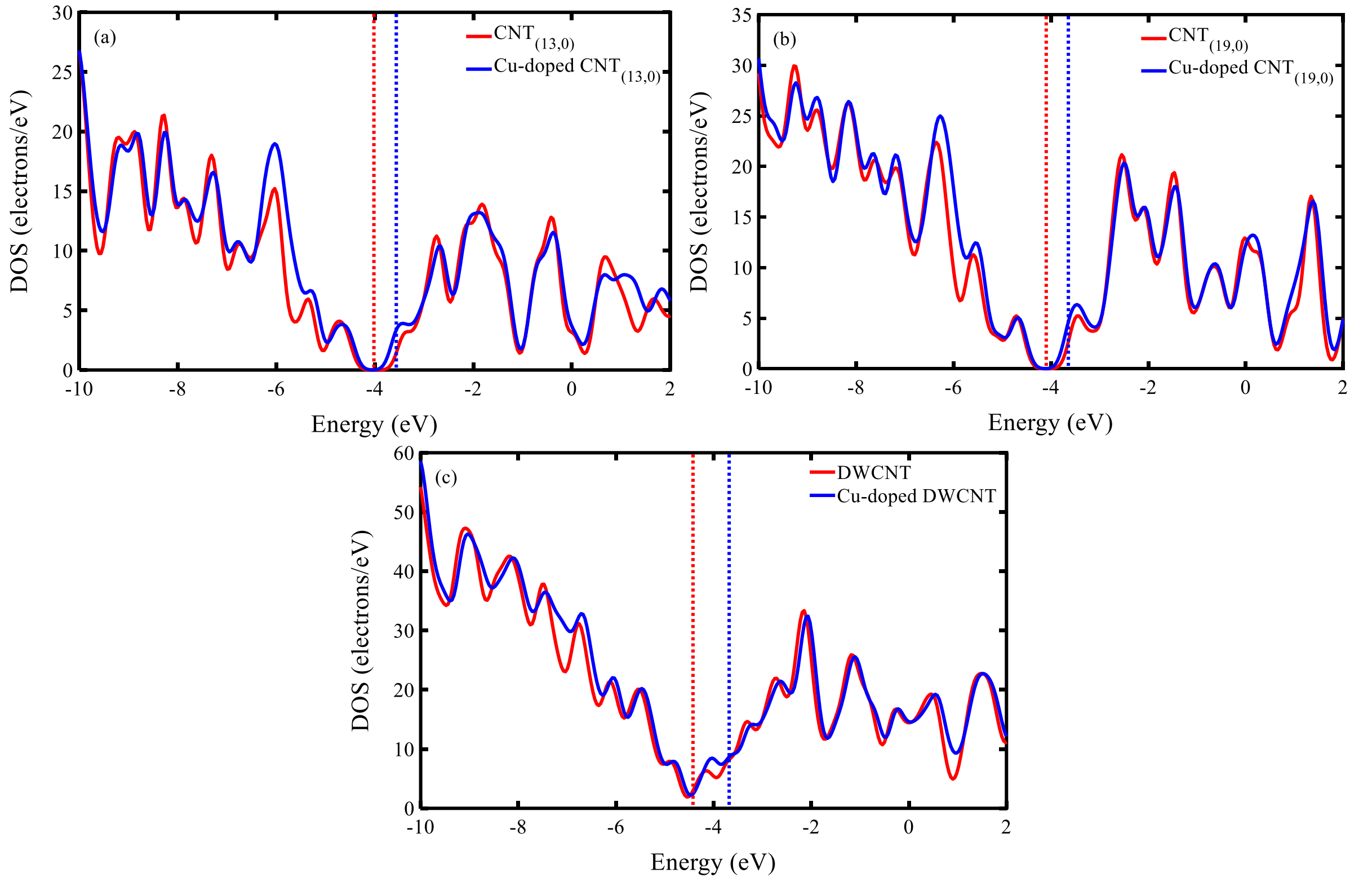

*Figure S2. DOS of pure and Cu-doped (a) $CNT_{(13,0)}$, (b) $CNT_{(19,0)}$, and (c) DWCNT. Dashed lines indicate the Fermi Energy ($E_F$) per configuration.*

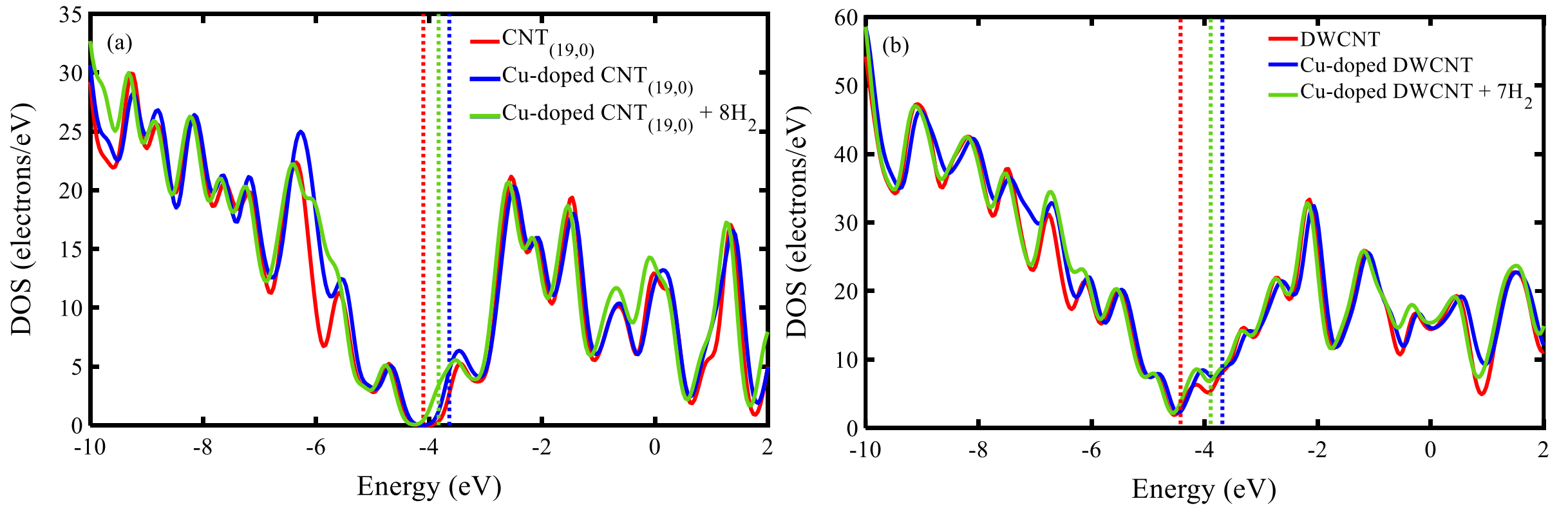

*Figure S3. DOS of pure, Cu-doped, and Cu-doped (a) $CNT_{(19,0)}$, and (c) DWCNT after adsorbing eight and seven hydrogen molecules, respectively. Dashed lines indicate the Fermi Energy ($E_F$) per configuration.*

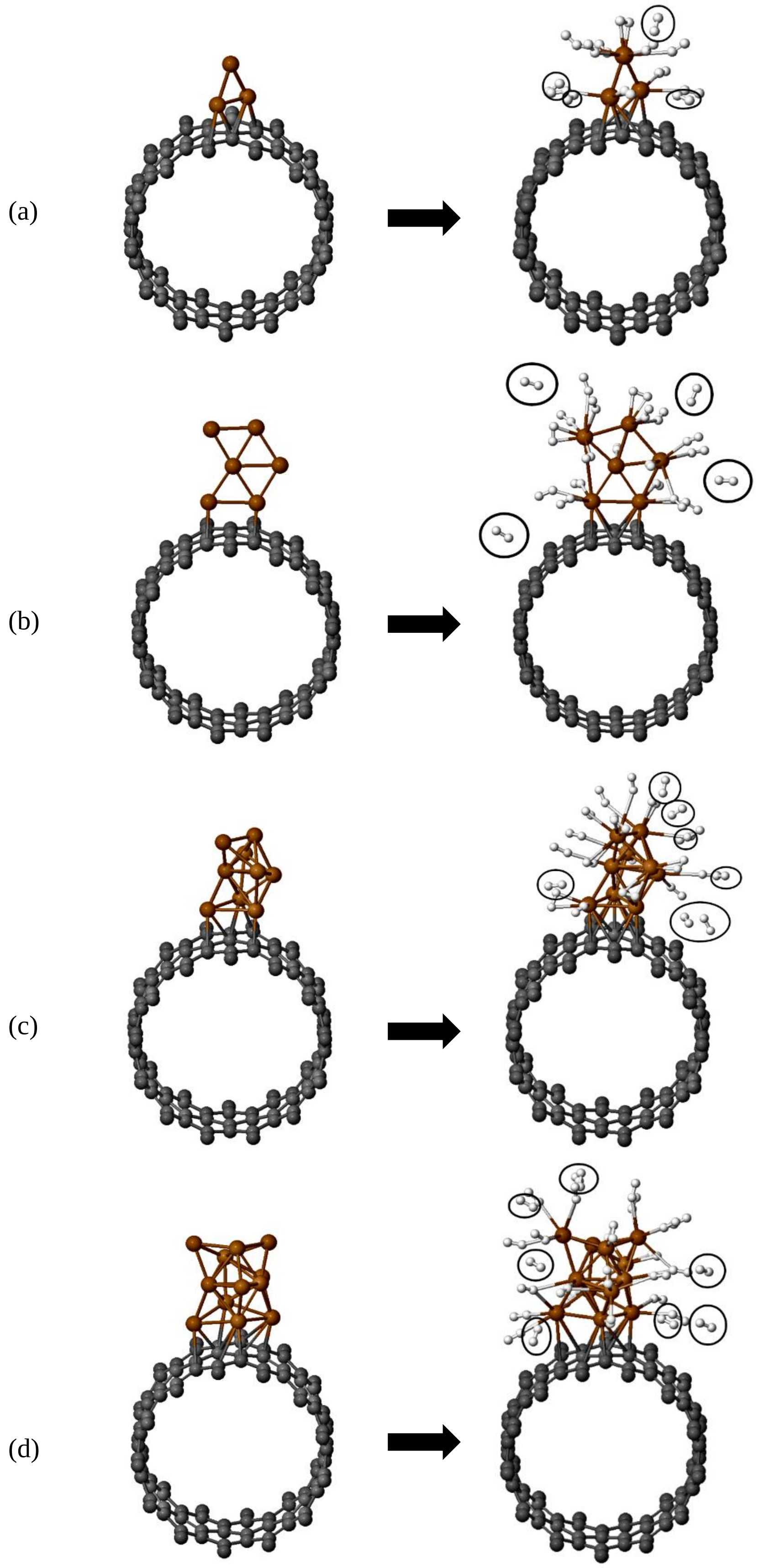

*Figure S4. Schematic representation of cluster formation and hydrogen adsorption on $CNT_{(13,0)}$ with cluster size of (a) $Cu_3$, (b) $Cu_6$ , (c) $Cu_9$, and (d) $Cu_{12}$ .*

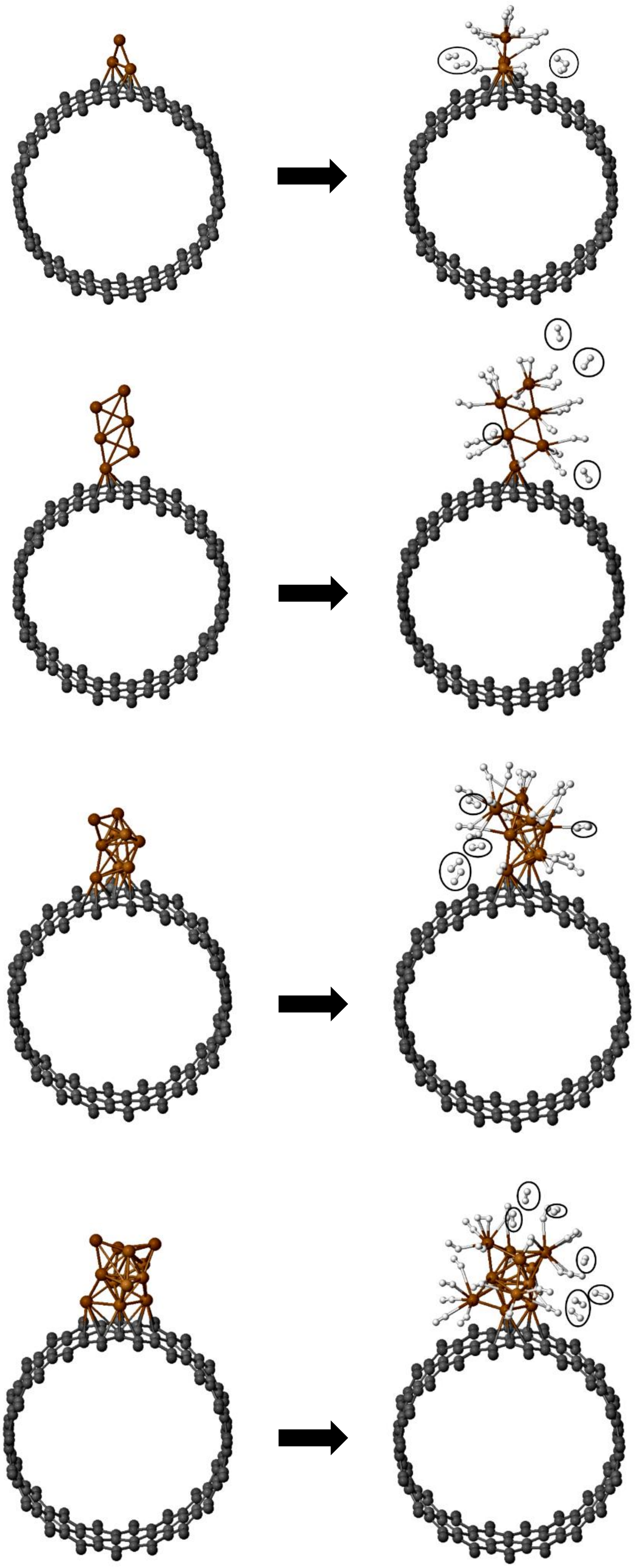

(a)

(b)

(c)

(d)

*Figure S5. Hydrogen adsorption on $CNT_{(19,0)}$ including a copper cluster containing (a) $Cu_3$, (b) $Cu_6$, (c) $Cu_9$, and (d) $Cu_{12}$.*

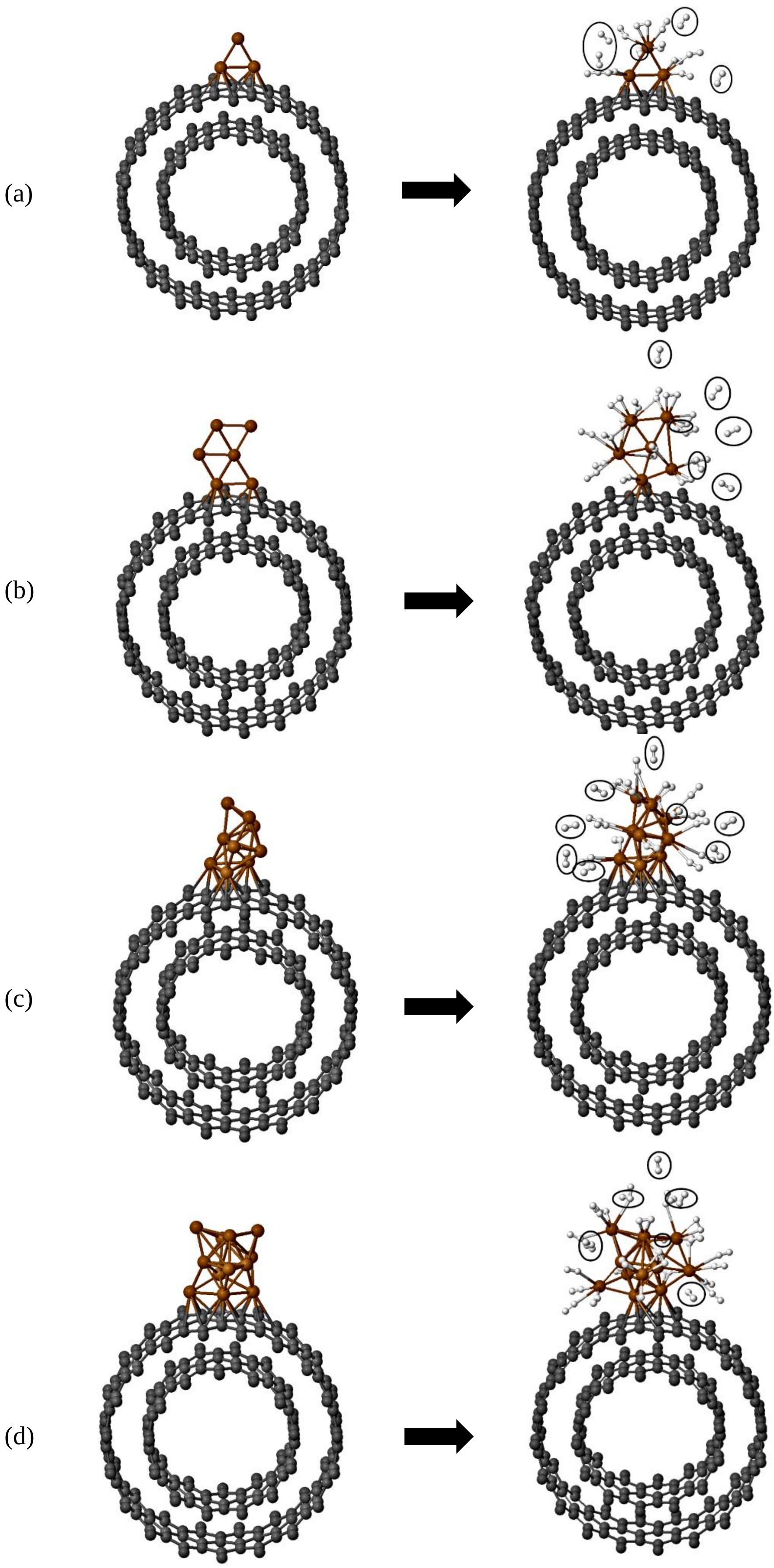

*Figure S6. Hydrogen adsorption on DWCNT including a copper cluster containing (a) $Cu_3$, (b) $Cu_6$, (c) $Cu_9$, and (d) $Cu_{12}$.*

*Table S3. Average Binding energies (eV) and Desorption temperatures between maximum number of adsorbed hydrogen molecules and the surface of $Cu_x$-doped single and double-walled CNTs.*

| **Cluster Size** | **Structure** | **$E_b$ $H_2$-($Cu_x$-CNTs)** | **$E_b$ $Cu_x$-CNTs (eV)** | **Maximum number of adsorbed $H_2$** | **Desorption Temperature (K)** |
|---|---|---|---|---|---|
| **$Cu_1$** | $Cu_1$-$CNT_{(13,0)}$ | 0.155 | 1.12 | 8 | 198 |
| | $Cu_1$-$CNT_{(19,0)}$ | 0.158 | 1.12 | 8 | 202 |
| | $Cu_1$-DWCNT | 0.149 | 1.22 | 7 | 191 |
| **$Cu_3$** | $Cu_3$-$CNT_{(13,0)}$ | 0.191 | 1.86 | 10 | 244 |
| | $Cu_3$-$CNT_{(19,0)}$ | 0.195 | 1.78 | 10 | 250 |
| | $Cu_3$-DWCNT | 0.212 | 2.06 | 9 | 271 |
| **$Cu_6$** | $Cu_6$-$CNT_{(13,0)}$ | 0.190 | 1.41 | 18 | 243 |
| | $Cu_6$-$CNT_{(19,0)}$ | 0.193 | 1.35 | 18 | 247 |
| | $Cu_6$-DWCNT | 0.240 | 1.69 | 16 | 307 |
| **$Cu_9$** | $Cu_9$-$CNT_{(13,0)}$ | 0.571 | 1.64 | 16 | 729 |
| | $Cu_9$-$CNT_{(19,0)}$ | 0.261 | 1.23 | 19 | 333 |
| | $Cu_9$-DWCNT | 0.500 | 1.90 | 16 | 640 |
| **$Cu_{12}$** | $Cu_{12}$-$CNT_{(13,0)}$ | 0.372 | 1.07 | 22 | 473 |
| | $Cu_{12}$-$CNT_{(19,0)}$ | 0.480 | 1.06 | 24 | 614 |
| | $Cu_{12}$-DWCNT | 0.450 | 1.77 | 22 | 576 |

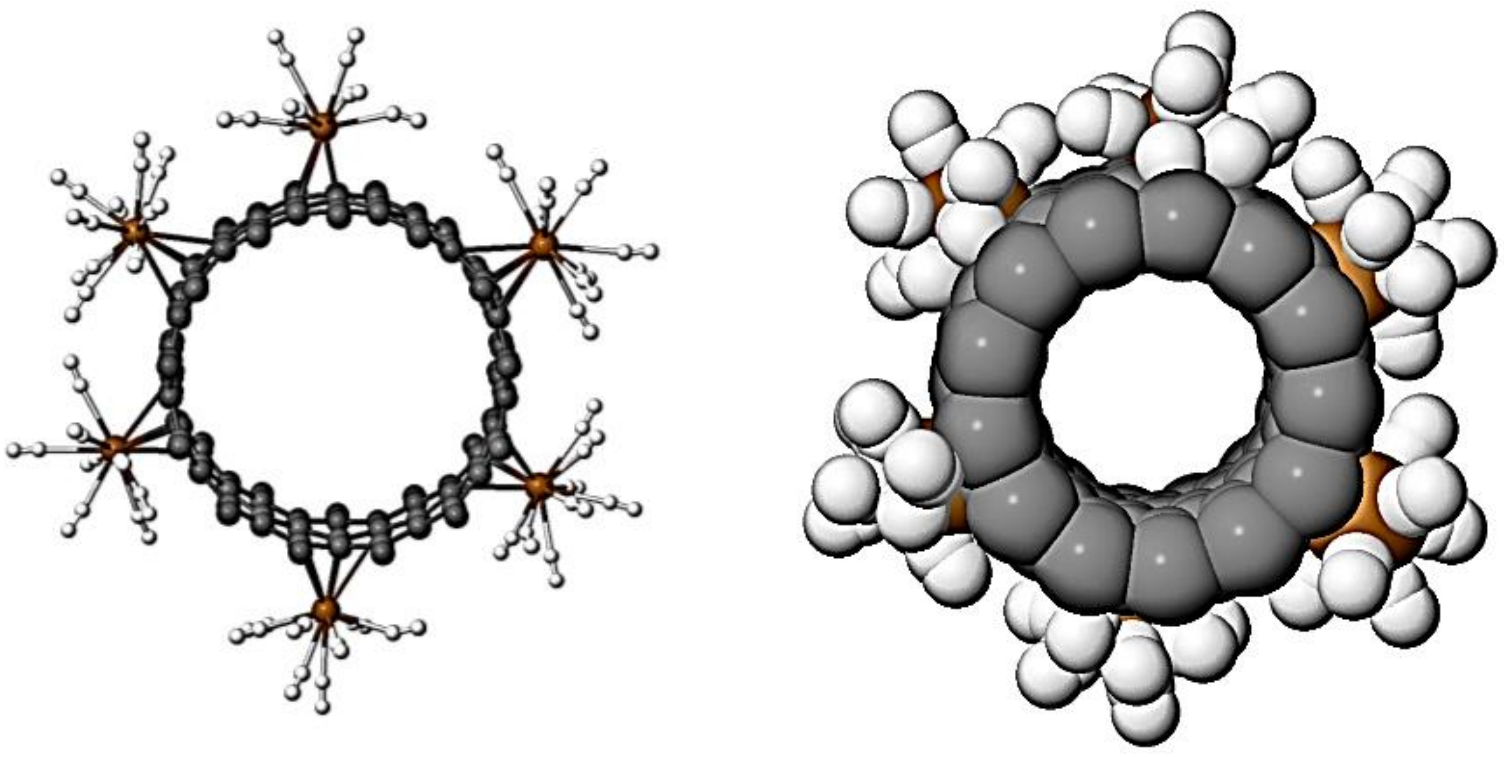

*Figure S7. Hydrogen adsorption on 6Cu-doped $CNT_{(13,0)}$.*

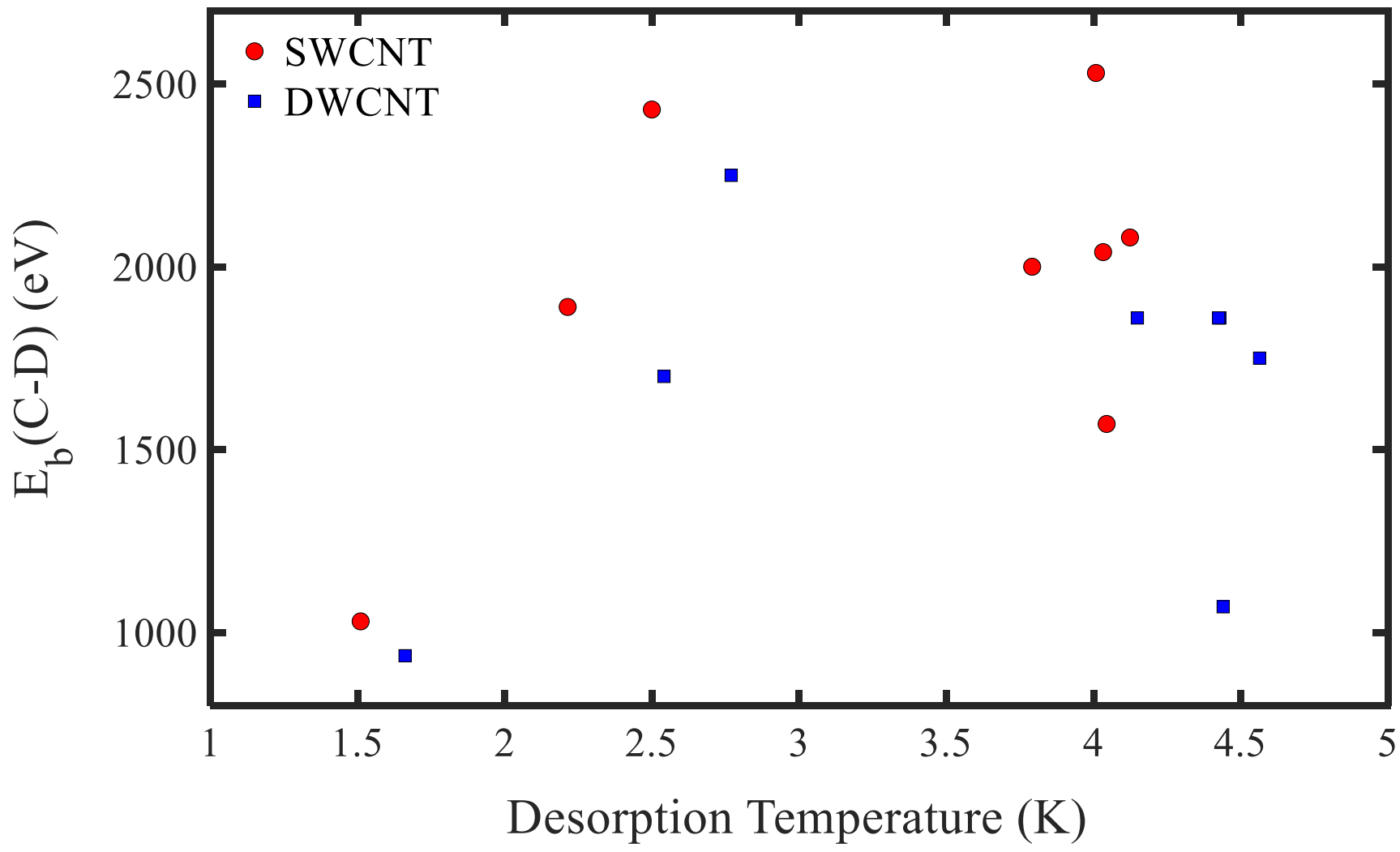

*Figure S8. Binding Energy between different impurities and the surface of single and double-walled CNT vs Desorption temperature.*